\documentclass[journal]{IEEEtran}

\usepackage{cite}

\usepackage[dvips]{graphicx} 

\ifCLASSOPTIONcompsoc
\usepackage[caption=false,font=normalsize,labelfon
t=sf,textfont=sf]{subfig}
\else
\usepackage[caption=false,font=footnotesize]{subfi
g}
\fi
\usepackage[cmex10]{amsmath}
\usepackage{amsthm,amssymb}
\usepackage{algorithm,algorithmic}

\begin{document}
%
\title{Smart Contract-Based Access Control for the Internet of Things}
%
%
%
%
\author{Yuanyu Zhang, \IEEEmembership{Member, IEEE},
		Shoji Kasahara, \IEEEmembership{Member, IEEE},
		Yulong Shen, \IEEEmembership{Member, IEEE},
		Xiaohong Jiang, \IEEEmembership{Senior Member, IEEE},
		and Jianxiong Wan  
\thanks{Y. Zhang and S. Kasahara are with the Graduate School of Information Science,
        Nara Institute of Science and Technology, Ikoma, Nara, Japan. E-mail:\{yyzhang,kasahara\}@is.naist.jp.}
\thanks{Y. Shen is with the School of Computer Science and Technology, Xidian University, Xi'an, Shannxi, China. E-mail: ylshen@mail.xidian.edu.cn.}
\thanks{X. Jiang is with the School of Systems Information Science, Future University Hakodate, Hakodate, Hokkaido, Japan, and the School of Computer Science and Technology, Xidian University, Shaanxi, China. E-mail: jiang@fun.ac.jp.}
\thanks{J.Wan is with the School of Data Science and Application, Inner Mongolia University of Technology, P.R. China. E-mail: jxwan@imut.edu.cn.}		
}

\maketitle

\begin{abstract}
This paper investigates a critical access control issue in the Internet of Things (IoT). In particular, we propose a smart contract-based framework, which consists of multiple access control contracts (ACCs), one judge contract (JC) and one register contract (RC), to achieve distributed and trustworthy access control for IoT systems. Each ACC provides one access control method for a subject-object pair, and implements both static access right validation based on predefined policies and dynamic access right validation by checking the behavior of the subject. The JC implements a misbehavior-judging method to facilitate the dynamic validation of the ACCs by receiving misbehavior reports from the ACCs, judging the misbehavior and returning the corresponding penalty. The RC registers the information of the access control and misbehavior-judging methods as well as their smart contracts, and also provides functions (e.g., register, update and delete) to manage these methods. To demonstrate the application of the framework, we provide a case study in an IoT system with one desktop computer, one laptop and two Raspberry Pi single-board computers, where the ACCs, JC and RC are implemented based on the Ethereum smart contract platform to achieve the access control.
\end{abstract} 

\IEEEpeerreviewmaketitle

\section {Introduction}\label{sec_into}
\IEEEPARstart{T}{hanks} to the rapid advance of communication and networking technologies (e.g., Wi-Fi, Zigbee, Bluetooth), a growing number of objects (e.g., sensors, actuators, smart devices) are being connected to the Internet nowadays, leading to the concept of the Internet of things (IoT) \cite{IoTYaqoob2017,Palattella2016JSAC}. The ubiquitous interconnection of physical objects significantly accelerates data collection, aggregation and sharing in the IoT, making the IoT one of the most fundamental architectures for various promising applications such as smart healthcare, intelligent transportation, home automation, etc. \cite{Xu2014TII,Al-Fugaha2015}. However, such interconnection may also incur crucial security issues into IoT systems, because adversaries can intrude into the systems to gain illegal access to the provided resources (e.g., data, services, storage units, computing units) by simply deploying their own or compromising existing IoT devices \cite{Orazio2017IoT, Bertino2017}. Thus, access control, which aims to prevent the illegal resource access from unauthorized entities, has been regarded as an increasingly vital research issue in the IoT for both academia and industry \cite{sicari2015security, Singh2016IoT,OUADDAH2017ComNet}.

Traditional IoT access control schemes are mainly built on top of the well-known access control models including the role-based access control model (RBAC) \cite{sandhu1996role}, the attributed-based access control model (ABAC) \cite{hu2015attribute} and the capability-based access control model (CapBAC) \cite{sandhu1994access}. In the RBAC-based schemes, the access control is based on the roles (e.g., administer, guest) of subjects (i.e., entities that access resources) within an organization. By associating the roles with access rights (e.g., read, write, execute) and assigning the roles to the subjects, the RBAC-based schemes can establish a many-to-many relationship between the access rights and the subjects \cite{Yavari2017ICDCS, Liu2017access}. The ABAC-based schemes implement the access control based on policies, which combine various types of attributes, such as subject attributes, object (i.e., the entity that holds resources) attributes and environment attributes, etc., to define a set of rules expressing under what conditions access rights can be granted to subjects \cite{ye2014efficient, bhatt2017NSS}. In the CapBAC-based schemes, access rights are granted to subjects based on the concept of capability, which is a transferable and unforgeable token of authority (e.g, a key, a ticket), and describes a set of access rights for each subject \cite{gusmeroli2013capability, mahalle2013identity}. 

It is notable that, in the above schemes, validating the access rights of subjects is usually conducted by a centralized entity, which turns out to be a single point of failure. To address this issue, distributed CapBAC models have been proposed recently \cite{Ramos2015JSAC, Hussein2017Commag}, where the access right validation is performed by the requested IoT objects themselves rather than a centralized entity. However, IoT objects are usually with low capability and thus may be easily compromised by adversaries, so they cannot be fully trusted as the access right validation entities. As a result, the distributed CapBAC models may fail to tackle the access control problem in untrustworthy IoT environments. Thus, a crucial question arises: how can we achieve distributed and trustworthy access control in the IoT? The answer may lie in the emerging blockchain technology, the key enabler behind modern cryptocurrency systems like the Bitcoin \cite{bitcoin} and Ethereum \cite{ethereum}. The blockchain is initially created as a distributed and immutable ledger of transactions for cryptocurrency systems. Thanks to the invention of smart contracts (executable codes that reside in the blockchain), the blockchain has now evolved into a promising platform for developing distributed and trustworthy applications, and has attracted considerable attentions from researchers in the IoT community \cite{Christidis2016Access, Kshetri2017IT}. Therefore, this paper aims to apply the smart contract-enabled blockchain technology to achieve distributed and trustworthy access control for the IoT.

Some initial work has been done on the blockchain-based access control. The authors in \cite{Dorri2017Percom} considered the access control issue in an IoT network with service providers, cloud storage, user devices and smart homes, each containing a miner and multiple IoT devices. Each home miner maintains a local private blockchain with a policy header storing access control policies to control all the access requests related to the home, i.e., internal, incoming and outgoing  requests. However, the authors eliminated the critical proof-of-work process \cite{nakamoto2008bitcoin} in the blockchain technology, resulting in an untrustworthy access control scheme. Notice that the main purpose of the blockchain in \cite{Dorri2017Percom} is to serve as a distributed and immutable storage for access control policies, whereas the computing capability of the blockchain was largely wasted. The idea of using the blockchain to only store access control policies has also been adopted in \cite{Zyskind2015, maesa2017blockchain}. Recently, the computing capability of the blockchain has been exploited in \cite{ouaddah2016fairaccess} for access control, where the blockchain plays the role of a decentralized access control manager. The authors used access tokens to represent access rights and the tokens can be delivered from one peer to another through transactions. When delivering a token, the sender embeds access control policies into the locking scripts of the transaction output. The receiver of the token must unlock the locking scripts to prove the possession of the token (i.e., the access rights to a certain resource). Using this scheme, a peer can be granted access rights by receiving a token, grant access rights to another subject by delivering a token, and access an object by spending a token. Although using locking scripts for access control  is an excellent idea, the computing capability of locking scripts is significantly limited. Different from \cite{ouaddah2016fairaccess}, this paper utilizes smart contracts to provide a much higher computing capability for achieving various access control methods. Notice that the idea of using smart contracts for access control has been adopted in \cite{azaria2016medrec, ramachandran2017using}, where, different from this paper, the main purpose of the smart contracts is to manage data records. 

To address the limitations of the above works, this paper proposes a smart contract-based access control framework, which consists of multiple access control contracts (ACCs), one judge contract (JC) and one register contract (RC), to achieve distributed and trustworthy access control for IoT systems. In the framework, each ACC provides one access control method for a subject-object pair, which implements both static access right validation based on predefined access control policies and dynamic access right validation by checking the behavior of the subject. The ACCs also provide functions for adding, updating and deleting access control policies. Once called by a subject for access control, the ACC will be run and verified by most participants in the system,  ensuring  the trustworthiness of the access control. To facilitate the dynamic validation of the ACCs, the JC provides a misbehavior-judging method, which receives misbehavior reports about the subjects from the ACCs, judges the misbehavior and returns the corresponding penalty. To manage the access control and misbehavior-judging methods, the RC registers the information (e.g., name, subject, object, smart contract) of the methods and also provides functions to register a new method and update or delete an existing method. To demonstrate the application of the framework, we provide a case study, in which we employ the Ethereum smart contract platform to implement the ACCs, JC and RC for the access control in a IoT system with one desktop computer, one laptop and two Raspberry Pi single-board computers.

The remainder of this paper is organized as follows. Section \ref{sec_sys} presents the IoT system considered in this paper and Section \ref{sec_plat} introduces the underlying smart contract platform for our access control framework. We introduce the distributed smart contract-based framework in Section \ref{sec_framework} and provide a case study for the proposed framework in Section \ref{sec_case}. Finally, Section \ref{sec_con} concludes this paper.

\section{System Architecture}\label{sec_sys} 

\begin{figure}[!t]
\centering
\includegraphics[width=3in]{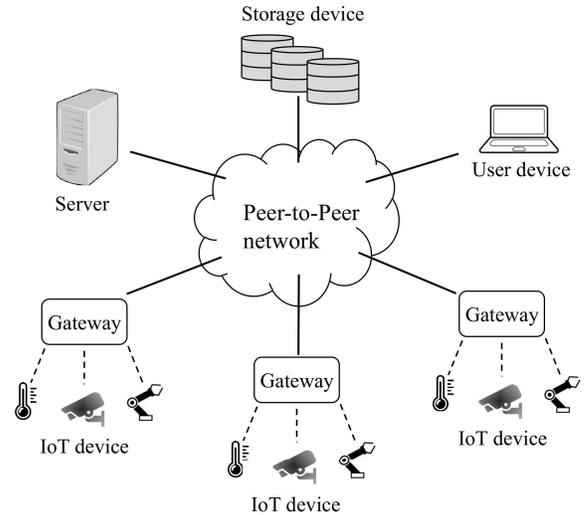}
\caption{Illustration of the considered IoT system.}
\label{fig_sysmodel}
\end{figure}

As illustrated in Fig. \ref{fig_sysmodel}, the IoT system considered in this paper consists of a large number of servers, storage devices, IoT gateways and user devices, which are connected together through a peer-to-peer (P2P) network. Also present in the system are numerous IoT devices (e.g.,  sensors and actuators), which are connected to the P2P network via the IoT gateways. The main roles of the peers are explained as follows.

\begin{itemize}
\item \textbf{Server}: A server is a device or a cluster of devices that can interact with the IoT devices and storage devices to provide a variety of services (e.g., smart home) for users. Interactions between the servers and other peers (e.g., IoT devices, storage devices) include collecting environmental data from the sensors, sending commands to the actuators to perform some operation, querying data from or storing data to the storage devices, etc. 

\item \textbf{Storage device}: A storage device can store data for other peers of the system, like the servers, sensors and users. Various data can be stored on the storage devices, like the application data of the servers, environmental data gathered by the sensors, user profiles, etc.

\item \textbf{User device}: A user device is a device (e.g., PCs, laptops, smart phones) through which users can enjoy the services (e.g., checking the current temperature of his/her own house) provided by the servers and read data from or write data to the storage devices.

\item \textbf{IoT gateway}: Each IoT gateway connects a cluster of IoT devices to the P2P network via short-range communication technologies like Bluetooth, Wi-Fi and Zigbee, and serves as the service agent for these IoT devices at the same time.

\item \textbf{IoT device}: The IoT devices in the system mainly include sensors, which can perceive environmental data (e.g., temperature) and send these data to the servers or storage devices for further use, and actuators, which can perform some operations (e.g., turning on the air conditioner) once receiving a command from users.
\end{itemize}
In typical IoT applications, each peer may have some resources (e.g., services, data, storage space) that are needed by the other peers. Thus, access control must be implemented by all resource owners to prevent unauthorized use of their resources. For example, a server must be able to block the access requests from users who has not signed up, or the access requests from signed-up users for some services that they have not subscribed. To prevent illegal use of its storage space and data, a storage device must be able to restrict the access requests from unauthorized peers for querying data or storing data. An IoT device must be able to deny the unauthorized access requests for retrieving its data or controlling its actuators. 

The aim of this paper is to address the critical access control issue for the above IoT system. In particular, we will propose an access control framework based on smart contracts to implement distributed and trustworthy access control. 

\section{Smart Contract Platform}\label{sec_plat}

\subsection{Ethereum Platform}
The proposed framework is based on the Ethereum smart contract platform \cite{ethereum}, through which each peer of the system can implement access control for its resources. The main elements of the Ethereum platform are briefly introduced as follows. For a detailed introduction to the Ethereum platform, please refer to \cite{ethereumIntro}.
\begin{figure}[!t]
\centering
\includegraphics[width=3.5in]{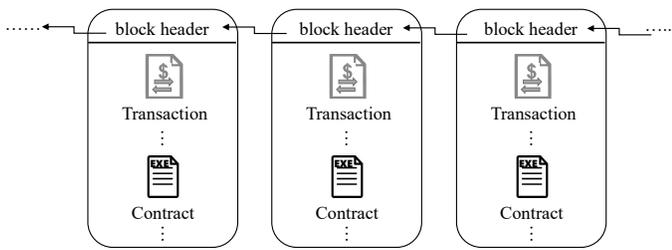}
\caption{Illustration of a blockchain.}
\label{fig_blockchain}
\end{figure}
\begin{itemize}
\item \textbf{Account/Address}: Ethereum has two types of accounts: externally controlled accounts and contract accounts, both identified by a 20-byte address. We refer to the former simply as \emph{accounts} and the latter as \emph{smart contracts} or \emph{contracts} throughout this paper.  

\item \textbf{Smart contract}: A smart contract or contract is regarded as a special account that has associated code (i.e., its functions) and data (i.e., its state) \cite{smartcontract}. In general, a smart contract is compiled into a piece of bytecode in an Ethereum-specific binary format (i.e., Ethereum Virtual Machine bytecode) and deployed by an account  to a global database known as blockchain. A smart contract usually provides many functions or application binary interfaces (ABIs) that can be used to interact with it. These ABIs can be executed by sending a \emph{transaction} from an account or a \emph{message} from another contract. They can also be executed by simply invoking the \emph{call} function without sending transactions and messages. Notice that only the former approach can modify the data (or state) of the contract.

\item \textbf{Transaction and Message}: A transaction is a data package signed and sent by an account to transfer some ether (Ethereum's native token) to another account. In addition to transferring ether, a transaction can also be sent with some parameters to execute the ABIs of a contract. A message is like a transaction, but it is sent by a contract instead of an account to run the associated ABIs of another contract.

\item \textbf{Blockchain}: Like most platforms such as Bitcoin, Ethereum also has a blockchain, which contains blocks of transactions and smart contracts with each block containing the hash of its previous block, as illustrated in Fig. \ref{fig_blockchain}. Every node connected to the network may have a local copy of the blockchain, and help maintain and update the blockchain by including new blocks.

\item \textbf{Mining}: Mining is a process that includes new blocks of transactions and contracts into the blockchain. The nodes performing this task are called miners. In one mining around, each miner constructs a block of newly generated transactions and contracts, and executes the proof-of-work consensus algorithm, where the miners repeatedly guesses random numbers to solve an extremely difficult cryptographic puzzle problem related to its block until one of them wins. The winning miner then broadcasts its block to the other nodes in the network to validate the block. For the block validation, each node not only checks the formats of the transactions and contracts in the block, but also executes the ABIs called by these transactions in its local EVM. If the formats of the transactions and contracts as well as the results of the called ABIs are valid, the other nodes will include the new block into its local blockchain; otherwise, they will discard the block. Through mining, the whole system reaches a common tamper-resistant consensus on the blockchain and no participant can deceive the others by wrongly executing the ABIs, as long as it controls no more than half of the computing power of the system. This is the key to achieving trustworthy access control for IoT systems. Notice that the mining in current implementation of the Ethereum is based on the concept of proof-of-work, while a novel proof-of-stake consensus algorithm \cite{proofofstake}, which depends on  the economic stake of miners instead of their computing computing power, will be used in future implementation of Ethereum.
\end{itemize}

\subsection{System Configurations}
To apply the Ethereum platform in our access control framework, we need to make the following basic configurations to the system.
\begin{itemize}
\item Each peer must be associated with an Ethereum account to represent itself in the system. Using the account, each peer can claim the deployment of a smart contract and identify itself during the access control. 
\item The Ethereum client can be run at all peers in the system except for IoT devices, due to the limited energy and computing power of IoT devices. All clients are assumed synchronized on the same block. Using the client, each peer except for IoT devices can directly interact with the blockchain to deploy smart contracts and send transactions to run the ABIs of smart contracts. These peers can also function as miners to conduct the mining task for the system.  
\item As IoT devices has no Ethereum clients, the IoT gateways act as agents for their local IoT devices to conduct access control for the resources of the IoT devices. To achieve this goal, each gateway stores the accounts of its local IoT devices and uses these accounts to sign transactions for deploying and running smart contracts on behalf of its local IoT devices. We assume gateways are physically accessible and thus unlikely to be compromised, so they can be trusted as the agents. 
\end{itemize}

\section{Access Control Framework}\label{sec_framework}
This section presents the smart contract-based distributed access control framework. We first introduce the system of smart contracts in the framework and then explain the main functions provided by the framework.

\begin{figure}[!t]
\centering
\includegraphics[width=3in]{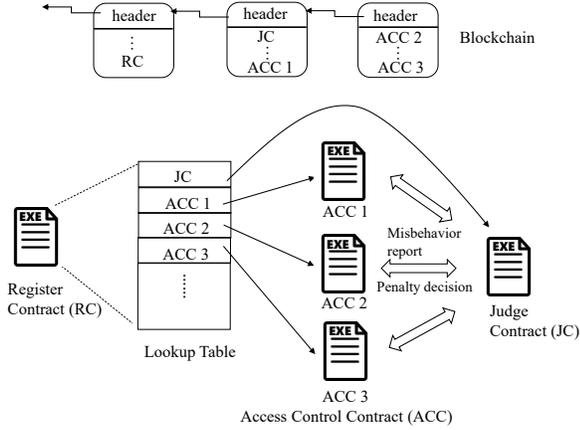}
\caption{Illustration of smart contract system.}
\label{fig_contractsys}
\end{figure}

\subsection{Smart Contract System} \label{sec_scs}
As illustrated in Fig. \ref{fig_contractsys}, the proposed framework consists of multiple access control contracts (ACCs), each of which implements the access control for a pair of peers, one judge contract (JC), which receives the misbehavior report of a peer from an ACC, judges the misbehavior and determines the corresponding penalty, and one register contract (RC), which stores the information of the JC and ACCs and provide functions to manage these contracts. Each of the contracts is introduced as follows.

\subsubsection{\textbf{Access control contract (ACC)}}\label{sec_ACC} An ACC (e.g., ACC 1, ACC 2, ACC 3 in Fig. \ref{fig_contractsys}) is deployed by a peer (object) who wants to control the access requests from another peer (subject). We assume that the subject-object pair can agree on multiple access control methods, and each method is implemented by one ACC. As a result, one subject-object pair can be associated with multiple ACCs, but one ACC can be associated with one and only one subject-object pair. In this framework, to control the access requests from the subject, each ACC implements not only static access right validation by checking predefined policies but also dynamic validation by checking the behavior of the subject. 

\begin{table}[t]
\renewcommand{\arraystretch}{1.2}
\caption{Illustration of policy list.}
\label{tb_accl}
\centering
\begin{tabular}{|l|l|l|l|}
\hline
\bfseries Resource &  \bfseries Action &\bfseries Permission &\bfseries ToLR  \\
\hline
file A  & read & allow & 2017-12-11 16:19 \\ 
\hline
file A  & write & deny & 2017-12-12 20:34 \\ 
\hline
Program A  & execute & deny & 2017-12-11 16:19 \\
\hline
...  & ... & ... & ... \\
\hline
\end{tabular}
\end{table}
An example of the ACC is given as follows. In this example, to achieve the access control, the ACC maintains a policy list as illustrated in Table \ref{tb_accl}, in which each row corresponds to the policy defined on a certain (resource, action) pair. The basic fields of each row are:
\begin{itemize}
\item \emph{Resource}: the resource for which the policy is defined, such as a file, a computing unit and a storage unit, etc.; 
\item \emph{Action}: the action that is performed on the resource, such as read, write, execute, etc.;
\item \emph{Permission}: the static permission predefined on the action, such as allow, deny, etc.; and 
\item \emph{Time of last request (ToLR)}: the time of the last access request from the subject.  
\end{itemize}
The \emph{Permission} field can be used for static validation and the \emph{ToLR} can be used for dynamic validation, such as detecting the misbehavior that the subject sends access requests too frequently in a short period of time. 

To record the misbehavior that the subject has exhibited on a certain resource as well as the corresponding penalty, the ACC also maintains a misbehavior list for each resource (as illustrated in Table \ref{tb_msbl}), where each row has the following basic fields:
\begin{itemize}
\item \emph{Misbehavior}: the misbehavior of the subject on this resource, such as too frequent request in a short period of time, etc.; 
\item \emph{Time}: the time when the misbehavior is exhibited; and 
\item \emph{Penalty}: the penalty on the subject for its misbehavior, such as blocking its access requests for a certain period of time, etc.;
\end{itemize}
The \emph{Misbehavior} field may also describe the details of the misbehavior to facilitate the misbehavior judging at the JC. 
\begin{table}[t]
\renewcommand{\arraystretch}{1.2}
\caption{Illustration of misbehavior list for each resource.}
\label{tb_msbl}
\centering
\begin{tabular}{|l|l|l|}
\hline
\bfseries Misbehavior &  \bfseries Time &\bfseries Penalty  \\
\hline
Too frequent access  & 2017-12-11 16:19 & blocked for 2 hours\\ 
\hline
Too frequent access  & 2017-12-12 20:34 & blocked for 4 hours\\ 
\hline
...  & ... & ... \\
\hline
\end{tabular}
\end{table}

The ACC also provides the following main ABIs to manage the policies and implement the access control.
\begin{itemize}
\item \emph{policyAdd()}: This ABI receives the information of a new access control policy and adds the information to the policy list. 
\item \emph{policyUpdate()}: This ABI receives the information of a policy that needs to be updated and updates the policy.  
\item \emph{policyDelete()}: This ABI receives the identification information of a policy and deletes the policy. 
\item \emph{accessControl()}: This ABI receives the information required for access control and returns the access result and penalty.  This ABI implements both the static and dynamic validation. When the subject calls (by sending a transaction) this ABI to authorize its current access request, both the static and dynamic validation processes will start to check the validity of the request. Once a possible misbehavior is detected, the ACC reports it to the JC by sending a message to execute the \emph{misbehaviorJudge} ABI of the JC, receives a penalty decision on the misbehavior from the JC and takes countermeasures based on the penalty decision. The access request is authorized if and only if it successfully passes both the static and dynamic validation processes.
\item \emph{setJC()}: In order for the ACC to execute the ABI of the JC, the ACC needs to keep an instance of the JC, so this ABI is to receive the address of the JC and set the JC instance.
\item \emph{deleteACC()}: This ABI performs the \emph{selfdestruct} operation to remove the code and storage of the ACC from the blockchain \cite{smartcontract}, such that the ACC can no longer be available. 
\end{itemize}
Notice that only the creator of the ACC can add a new policy, update or delete an existing policy, set the JC and delete the ACC. Thus, permission must be carefully considered in the implementation of the ABIs.
 
\subsubsection{\textbf{Judge contract (JC)}} 
\begin{figure}[!t]
\centering
\includegraphics[width=3in]{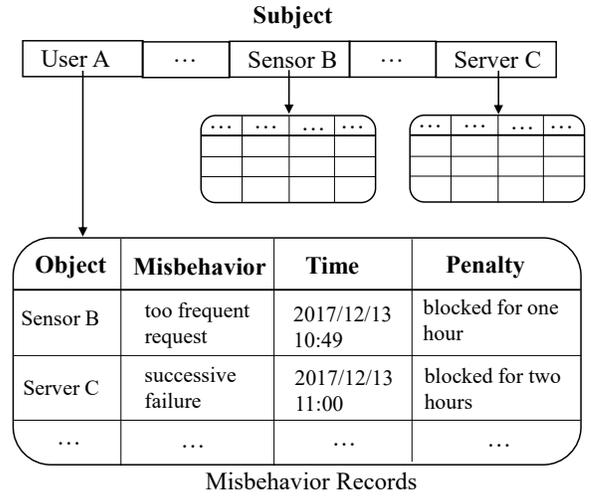}
\caption{Illustration of misbehavior records.}
\label{fig_misbehavior}
\end{figure}

The JC implements a misbehavior-judging method, which judges the misbehavior of the subject and determines the corresponding penalty, when receiving a potential misbehavior report from an ACC, as illustrated in Fig. \ref{fig_contractsys}. The penalty can be based on the misbehavior history of the subject, so the JC may need to keep a record of the misbehavior history of all subjects. After determining the penalty, the JC returns the decision to the ACC for further operation. Here, we give an example of the JC, which maintains a misbehavior list for each subject who has behaved abnormally, as illustrated in Fig. \ref{fig_misbehavior}. The fields of each record include:
\begin{itemize}
\item \emph{Object}:  the peer who suffered from the misbehavior;
\item \emph{Misbehavior}: the details of the misbehavior; 
\item \emph{Time}: the time when the misbehavior is exhibited; and
\item \emph{Penalty}: the penalty imposed on the misbehavior.
\end{itemize}

The JC also provides the following main ABIs for judging misbehavior, determining the penalty and managing the JC.
\begin{itemize}
\item \emph{misbehaviorJudge()}: This ABI can be run by any ACC to report the misbehavior of a subject to the JC. After receiving the report, this ABI judges the misbehavior of the subject, determines the penalty on the subject based on the misbehavior history of the subject and returns the penalty decision to the ACC that reported the misbehavior. This ABI also adds a new misbehavior record to the misbehavior list of the subject.
\item \emph{deleteJC()}: This ABI performs the \emph{selfdestruct} operation to delete the JC. 
\end{itemize}

\subsubsection{\textbf{Register contract (RC)}} The main role of the RC in the system is to manage the access control and misbehavior-judging methods. To achieve this goal, the RC maintains a lookup table, which registers the required information to find and execute all the methods. An example of the lookup table is given in Table \ref{tb_lt}, in which each row contains the following information of a method:  
\begin{itemize}
\item \emph{MethodName}: the name of the method;
\item \emph{Subject}: the subject of the corresponding subject-object pair of the method; 
\item \emph{Object}: the object of the corresponding subject-object pair of the method; 
\item \emph{ScName}: the name of the corresponding smart contract implementing this method;
\item \emph{Creator}: the peer who created and deployed the contract; 
\item \emph{ScAddress}: the address of the smart contract; and 
\item \emph{ABI}: the ABIs provided by the contract.
\end{itemize}
For the JC, the \emph{Subject} and \emph{Object} fields are left blank. In general, the object is the creator of the ACC as well as the creator of the access control method. Notice that for the case where the object is an IoT device, the creator is the local gateway, i.e., the agent for deploying contracts and sending transactions for the IoT device.

\begin{table*}[!t]
\renewcommand{\arraystretch}{1.2}
\caption{Illustration of lookup table.}
\label{tb_lt}
\centering
\begin{tabular}{|l|l|l|l|l|l|l|}
\hline
\bfseries MethodName &\bfseries Subject  &  \bfseries Object &\bfseries ScName &\bfseries Creator  &\bfseries ScAddress & \bfseries ABI\\
\hline
 Method 1&Server A & Sensor B & ACC 1&  Sensor B &0xca35b7d915458ef540ade6068dfe2f44e8fa733c  & accessControl(),...\\ 
\hline
 Method 2 & Server A & Sensor B & ACC 2& Sensor B &0xab072c469475346532bf47aea86df61761049565 &  accessControl(),...\\ 
\hline
 Method 3&Sensor B & Server A & ACC 3& Server A  &0xb51f6d86d4c998531056a501344060fbafc32a48 &  accessControl(),...\\
\hline
 JC &  &  & Judge &  & 0x3f23c7b929cced4191ef6064ffcb33902ea1d92b & misbehaviorJudge()...\\
\hline
...  & ... & ... & ...& ... &...&...\\
\hline
\end{tabular}
\end{table*}
With the help of the lookup table, the RC provides the following main ABIs to mange these methods.
\begin{itemize}
\item \emph{methodRegister()}: This ABI receives the information of a new method and registers the information into the lookup table. 
\item \emph{methodUpdate()}: This ABI receives the information of an existing method that needs to be updated and update the information, especially the fields of ScAddress and ABI.
\item \emph{methodDelete()}: This ABI receives the \emph{MethodName} of a method and deletes the method from the lookup table.
\item \emph{getContract()}: This ABI receives the \emph{MethodName} of a method and returns the address and ABIs of the contract (i.e., the ACCs and JC) of the method.
\end{itemize}
Notice that only the creator of the method can register, update and delete the method.  

\subsection{Main Functions of the Framework}
With the help of the ACC, JC and RC smart contracts, the framework can provide many functions to facilitate the access control of the IoT system. These functions mainly include registering, updating and deleting an access control method; registering and updating the misbehavior-judging method; adding, updating and deleting a policy of an ACC; and the access control for a subject-object pair. The process of each function is explained as follows.

\subsubsection {\textbf{Registering a new access control method}} A subject-object pair can agree on a new access control method, which is registered by the creator (i.e., the object) of the method through the following steps. 
\begin{itemize}
\item Step 1: Create (i.e., write and compile) an ACC for the new method.
\item Step 2: Send a transaction to deploy the newly created ACC onto the blockchain.  
\item Step 3: Send a transaction to call the \emph{methodRegister} ABI of the RC to register the required information of the new ACC in the lookup table of the RC.
\end {itemize}    
Registering the misbehavior-juding method follows the same steps as above.
\subsubsection{\textbf{Updating an existing access control method}} A subject-object pair can agree on updating an existing access control method, which is conducted by the creator of the method through the following steps.
\begin{itemize}
\item Step 1: Create a new ACC, which is used to replace the old one.
\item Step 2: Send a transaction to deploy the newly created ACC onto the blockchain.
\item Step 3: Send a transaction to run the \emph{methodUpdate} ABI of the RC to update the ACC-related fields of the method, such as the ScName, ScAddress, ABI, etc..
\item Step 4: Send a transaction to run the \emph{deleteACC} ABI of the old ACC to destruct it.
\end {itemize}  
Updating the misbehavior-juding method follows the same steps as above.
\subsubsection{\textbf{Deleting an existing access control method}} A subject-object pair can agree on deleting an existing access control method, which is conducted by the creator of the method through the following steps. 
\begin{itemize}
\item Step 1: Send a transaction to run the \emph{methodDelete} ABI of the RC to delete the information of the existing method from the lookup table.
\item Step 2: Send a transaction to run the \emph{deleteACC} ABI of the ACC of the method.
\end {itemize}  

\subsubsection{\textbf{Adding, updating and deleting a policy}} A subject-object can agree on adding an access control policy for a newly-deployed resource, which is conducted by the creator of the method through sending a transaction to call the \emph{policyAdd} ABI of the corresponding ACC. Similarly, the creator can send a transaction to call the \emph{policyUpdate} (resp. \emph{policyDelete}) ABI of the ACC to update (resp. delete) an existing policy of the access control method. 
\subsubsection{\textbf {Access control}} The ACC for the access control between a subject-object pair can be called by either the subject or the object. We assume that both the subject and object know the names of all the available methods for the access control between them. The illustration of the case where the ACC is called by the subject is given in Fig. \ref{fig_accessCon_subject}, where a server (the subject) wants to access the resource of an IoT device (the object). To complete the access control, the following steps are executed:

\begin{figure}[!t]
\centering
\subfloat[ACC called by the subject.]{\includegraphics[width=3in]{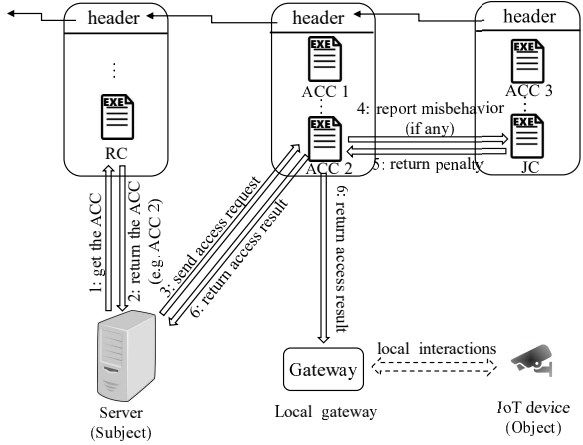}
\label{fig_accessCon_subject}}
\hfil
\subfloat[ACC called by the object.]{\includegraphics[width=3in]{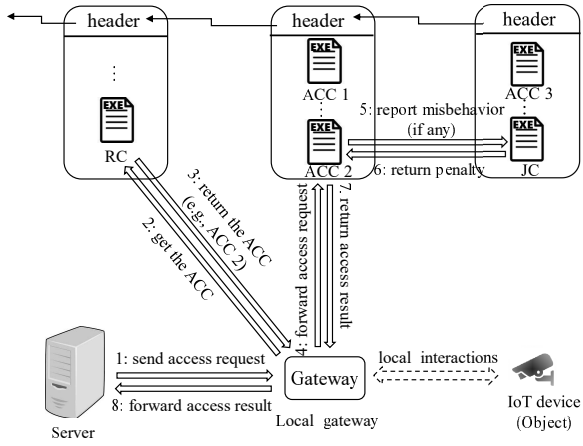}
\label{fig_accessCon_object}}
\caption{Illustration of access control.}
\label{fig_accessCon}
\end{figure}

\begin{itemize}
\item Step 1: The server calls the \emph{getContract} ABI of the RC to retrieve the ACC (e.g., the ACC 2 in Fig. \ref{fig_accessCon_subject}) for the access control. 
\item Step 2: The RC returns the address and ABI of the ACC to the server. 
\item Step 3: The server sends a transaction, which contains the required information for access control, to call the \emph{accessControl} ABI of the ACC. This transaction will be encapsulated in a new block and the \emph{accessControl} ABI will not be executed until the new block is mined and included in the blockchain by some miner. 
\item Step 4: During the access control process, the ACC will send a message to call the \emph{misbehaviorJudge} ABI of the JC, if some potential misbehavior of the subject is detected.
\item Step 5: Once the \emph{misbehaviorJudge} ABI completes juding the misbehavior and determining the penalty, it will return the penalty to the ACC. 
\item Step 6: Finally, the access result will be returned to both the subject and object, after the access control process finishes. 
\end{itemize}
Since all miners will reach a consensus on the result of the access control through mining, so no miners can tamper with the access control process. As the agent of the IoT device, the local gateway informs the IoT device the real-time status of the access control, such as the arrival of access requests and the access results, via secure local interactions. Fig. \ref{fig_accessCon_object} illustrates the case where the ACC is called by the object. The main difference between the access control in Fig. \ref{fig_accessCon_object}  and that in Fig. \ref{fig_accessCon_subject} is that the access request of the subject (resp. the access result) in Fig.  \ref{fig_accessCon_object} is forwarded by the object rather than being directly sent to the \emph{accessControl} ABI of the ACC (resp. the subject).  
\begin{table*}[!t]
\renewcommand{\arraystretch}{1.2}
\caption{Specifications of devices.}
\label{tb_spec}
\centering
\begin{tabular}{|l|l|l|l|l|}
\hline
\bfseries Device &\bfseries CPU  &  \bfseries Operating System &\bfseries Memory &\bfseries Hard Disk\\
\hline
 Dell Inspiron 3650 &Intel Core i7-6700, 3.40GHz & Windows 10 Home (64 bit) & 16GB& 2TB\\ 
\hline
 MacBook Pro & Intel Core i5, 2GHz & macOS Sierra (Version 10.12.6) & 8GB & 256GB\\ 
\hline
 Raspberry Pi 3 Model B &quad-core ARM Cortex A53, 1.2GHz & Raspbian GNU/Linux 8 (jessie) & 1GB SDRAM & 16GB (microSD card)\\
\hline
\end{tabular}
\end{table*}

\section{Case Study}\label{sec_case}
This section provides a case study to demonstrate the application of the proposed framework for distributed access control in the IoT. We first introduce the hardware and software used in the study and then present how the access control is implemented based on the framework. Finally, we show some experiment results.

\begin{figure}[!t]
\centering
\includegraphics[width=3in]{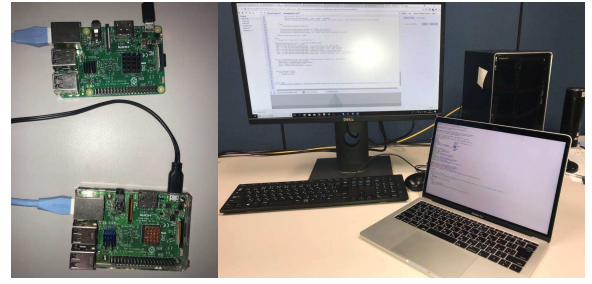}
\caption{Hardware used in the case study.}
\label{fig_hardware}
\end{figure}
\subsection{Hardware and Software}
We considered a case with one desktop computer (Dell Inspiron 3650), one laptop (MacBook Pro) and two single-board computers (Raspberriy Pi 3 Model B), as shown in Fig. \ref{fig_hardware}. The specifications of these devices are listed in Table \ref{tb_spec}. The desktop and laptop correspond to the user devices in the system and the single-board computers correspond to the local gateways. We considered the access control issue between the single-board computers, of which one serves as the subject (or the agent of the subject) and the other severs as the object (or the agent of the object). 

\begin{figure}[!t]
\centering
\includegraphics[width=3in]{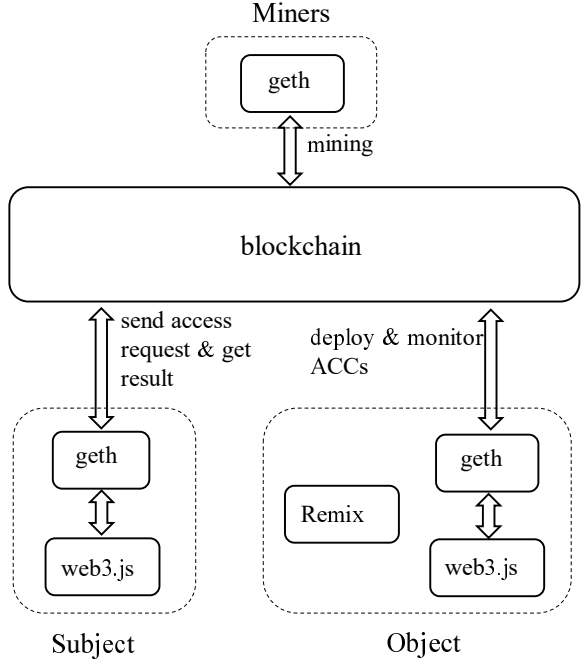}
\caption{Software used in the case study.}
\label{fig_softwares}
\end{figure}

On each device, a geth client \cite{geth} (a command line interface implemented in the Go language) was installed to transform the device into an Ethereum node. With the geth clients, we created an Ethereum account for each node and configured these nodes to form a private blockchain network (as illustrated in Fig. \ref{fig_softwares}), where the desktop computer and the laptop play the roles of miners due to their relatively large computing and storage capability, and the single-board computers function as lightweight Ethereum nodes that deploy ACCs and send transactions for access control.   

For writing and compiling the ACC at the object side, we utilized the Remix integrated development environment (IDE) \cite{remix}, which is a browser-based IDE for Solidity (i.e., the programming language for writing smart contracts) \cite{solidity}. In addition, we adopted the web3.js \cite{web3js} (i.e., the official Ethereum JavaScript API) at the object side to interact with the corresponding geth client through HTTP connections for deploying the compiled ACC and also monitoring the states of the ACC ( i.e., the results of the access control). The web3.js was also installed at the subject side to interact with the geth for sending access requests to the ACC via transactions and also receiving the access control results from the ACC. 

\subsection{Implementation}
The implementation of the ACC, RC and JC is based on the examples in Section \ref{sec_scs}. 
\subsubsection{ACC}
In this implementation, we defined a simple misbehavior, which is sending access requests too frequently in a short period of time. To help characterize the misbehavior, we added the following fields to the rows (i.e., policies) in Table \ref{tb_accl}:
\begin{itemize}  
\item \emph{minInterval}: the minimum allowable time interval between two successive requests. If the time interval between two successive requests is less than or equal to \emph{minInterval}, the later request will be treated as a frequent request.
\item \emph{NoFR}: the number of frequent requests in a short time period.
\item \emph{threshold}: the threshold on the \emph{NoFR}. If the \emph{NoFR} is larger than or equal to the \emph{threshold}, the ACC judges that a misbehavior occurs.
\end{itemize}
As the penalty for the misbehavior, the access requests from the subject will be blocked for a certain time period. We introduced a variable \emph{timeOfUnblock} for each resource to represent the time until when requests are blocked, which is set to $0$ when the requests are unblocked. We used a struct to store the fields of a policy and applied a two-dimensional mapping from the fields of \emph{resource} (primary key) and \emph{action} (secondary key) to this struct to construct the policy list. The ACC also contains a JC instance, through which the \emph{misbehaviorJudge} ABI of the JC can be run by the ACC. Based on the above fields and variables, we designed the \emph{accessControl} ABI as in Algorithm \ref{alg_acc}, which receives the inputs of \emph{resource}, \emph{action} and \emph{time} (i.e., when the request is sent), and returns the access \emph{result} and \emph{penalty}.
 \begin{algorithm}
 \caption{accessControl ABI}
 \label{alg_acc}
 \begin{algorithmic}[1]
 \renewcommand{\algorithmicrequire}{\textbf{Input:}}
 \renewcommand{\algorithmicensure}{\textbf{Output:}}
 \REQUIRE \emph{resource}, \emph{action}, \emph{time}
 \ENSURE  \emph{result}, \emph{penalty}
 \\ \textit{Require}: \emph{policyCheck} $\leftarrow$ false, \emph{behaviorCheck} $\leftarrow$ true, \emph{penalty} $\leftarrow$ 0, JC instance \emph{judge}, policy list \emph{policies}, \emph{timeofUnblock} of \emph{resource}.\\
  \IF {This request is from the subject}
	\STATE \emph{p} $\leftarrow$ \emph{policies}[\emph{resource}][\emph{action}].
		\IF {\emph{timeofUnblock} $\leq$ \emph{time}}
				\IF{\emph{timeofUnblock} $>$ 0}
				\STATE \emph{p.NoFR} $\leftarrow$ 0, \emph{p.ToLR} $\leftarrow$ 0, \emph{timeofUnblock} $\leftarrow$ 0.
				\ENDIF
				\IF{\emph{p.policy} = "allow"}
				\STATE \emph{policyCheck} $\leftarrow$ true.
				\ELSE
				\STATE \emph{policyCheck} $\leftarrow$ false.
				\ENDIF
				\IF{\emph{time} $-$ \emph{p.ToLR} $\leq$ \emph{p.minInterval}}
				\STATE \emph{p.NoFR} $\leftarrow$ \emph{p.NoFR} + 1.
					\IF {\emph{p.NoFR} $\geq$ \emph{p.threshold}}
					\STATE Detect a misbehavior \emph{msb}.
					\STATE \emph{behaviorCheck} $\leftarrow$ false.
					\STATE \emph{penalty} $\leftarrow$ \emph{judge.misbehaviorJudge(subject, msb)}.
					\STATE \emph{timeofUnblock} $\leftarrow$ \emph{time} $+$ \emph{penalty}.
					\STATE Push \emph{msb} into the misbehavior list of \emph{resource}.
					\ENDIF
				\ELSE
				\STATE \emph{p.NoFR} $\leftarrow$ 0.
				\ENDIF
		\ENDIF
		\STATE \emph{p.ToLR} $\leftarrow$ \emph{time}.
  \ENDIF
	\STATE \emph{result} $\leftarrow$ \emph{policyCheck} \AND \emph{behaviorCheck}.
  \STATE Trigger event \emph{returnResult(result, penalty)}. 
 \end{algorithmic} 
 \end{algorithm}
The static validation is from Line 7 to Line 11 and the dynamic validation is from Line 12 to Line 23. The event \emph{returnResult(result, penalty)} in Line 28 is used to return the access result and penalty to both the subjects and objects. For the detailed implementation of the ACC, please refer to \cite{Sc-IoT}.

\subsubsection{RC}
The key issue in the implementation of the RC is to construct the loopup table as shown in Table \ref{tb_lt}. Like the construction of policy list for the ACC, we used a struct to store the information of each method and applied a mapping from the field of \emph{MethodName} to this struct to construct the loopup table. 
\subsubsection{JC}
In the implementation of the JC, we used a dynamic array to store the misbehavior records of a subject. We considered a simple misbehavior judging method, which treats all potential misbehavior received from the ACC as misbehavior. When receiving a misbehavior report of a subject from the ACC, the \emph{misbehaviorReport} ABI pushes the misbehavior into the misbehavior record array of the subject and then uses the following function to determine the corresponding penalty:
\begin{equation}\label{eqn_penalty}
penalty = (base)^{\lfloor \ell/interval\rfloor},
\end{equation}
where $\ell$ is the number of misbehavior that the subject has exhibited (i.e., the length of the misbehavior record array of the subject), and $base$ and $interval$ are parameters that determine how the penalty changes with $\ell$. Notice that $base$ and $interval$ are initialized when the JC is deployed.

\subsubsection{JavaScripts at the subject and object}
The access control in this study is implemented based on the case in Fig. \ref{fig_accessCon_subject}, where the ACC is called by the subject and the result is returned to both sides. To implement the access control, we created two JavaScripts (one at the subject and the other at the object) using the web3.js to interact with the JC and ACC. As shown in Algorithm \ref{alg_ar}, the JavaScript at the subject side first retrieves the address \emph{addr}  and ABI \emph{abi} of the ACC from the RC (Line 1 - Line 3) and then sends a transaction that contains the access request information  (\emph{resource}, \emph{action}, \emph{time}) to run the \emph{accessControl} ABI of the ACC for access control (Line 4 - Line 5). Finally, the JavaScript watches the event \emph{returnResult()} returned from the \emph{accessControl} ABI to retrieve the access result (Line 6 - Line 11). 
 \begin{algorithm}
 \caption{Access Request JavaScript}
 \label{alg_ar}
 \begin{algorithmic}[1]
 \renewcommand{\algorithmicrequire}{\textbf{Input:}}
 \renewcommand{\algorithmicensure}{\textbf{Output:}}
 \REQUIRE \emph{resource}, \emph{action}, \emph{time}
 \ENSURE  \emph{result}, \emph{penalty}
  \STATE Create a RC instance \emph{register}.
	\STATE Specify the access control method name \emph{method}.
	\STATE (\emph{addr}, \emph{abi})$\leftarrow$ \emph{register.getContract(method)}.
	\STATE Create an ACC instance \emph{acc} with \emph{addr}, \emph{abi}.
	\STATE Send a transaction containing parameters (\emph{resource}, \emph{action}, \emph{time}) to the \emph{accessControl} ABI of \emph{acc}.
	\WHILE{ture}
	\IF{Event \emph{returnResult()} is captured}
	\STATE (\emph{result}, \emph{penalty}) $\leftarrow$ \emph{returnResult()}.
	\STATE break.
	\ENDIF
	\ENDWHILE
 \RETURN \emph{result}, \emph{penalty} 
 \end{algorithmic} 
 \end{algorithm}

The JavaScript at the object side is illustrated in Algorithm \ref{alg_am}), which uses the same statements (Line 1 - Line 3) to retrieve the address and ABI of the ACC from the RC and infinitely watches the \emph{returnResult()} events from the ACC to know who wants to access which resource at what time, and what the corresponding result and penalty are (Line 4 - Line 10). 
 \begin{algorithm}
 \caption{Access Monitor JavaScript}
 \label{alg_am}
 \begin{algorithmic}[1]
 \renewcommand{\algorithmicrequire}{\textbf{Input:}}
 \renewcommand{\algorithmicensure}{\textbf{Output:}}
  \STATE Create a RC instance \emph{register}.
	\STATE Specify the access control method name \emph{method}.
	\STATE (\emph{addr}, \emph{abi})$\leftarrow$ \emph{register.getContract(method)}.
	\STATE Create an ACC instance \emph{acc} with \emph{addr}, \emph{abi}.
	\WHILE{ture}
	\IF{Event \emph{returnResult()} is captured}
	\STATE (\emph{result}, \emph{penalty}) $\leftarrow$ \emph{returnResult()}.
	\STATE Display \emph{result}, \emph{penalty}.
	\ENDIF
	\ENDWHILE
 \end{algorithmic} 
 \end{algorithm}

\subsection{Experiments}
\begin{figure}[!t]
\centering
\subfloat[Results at the object.]{\includegraphics[width=3.5in]{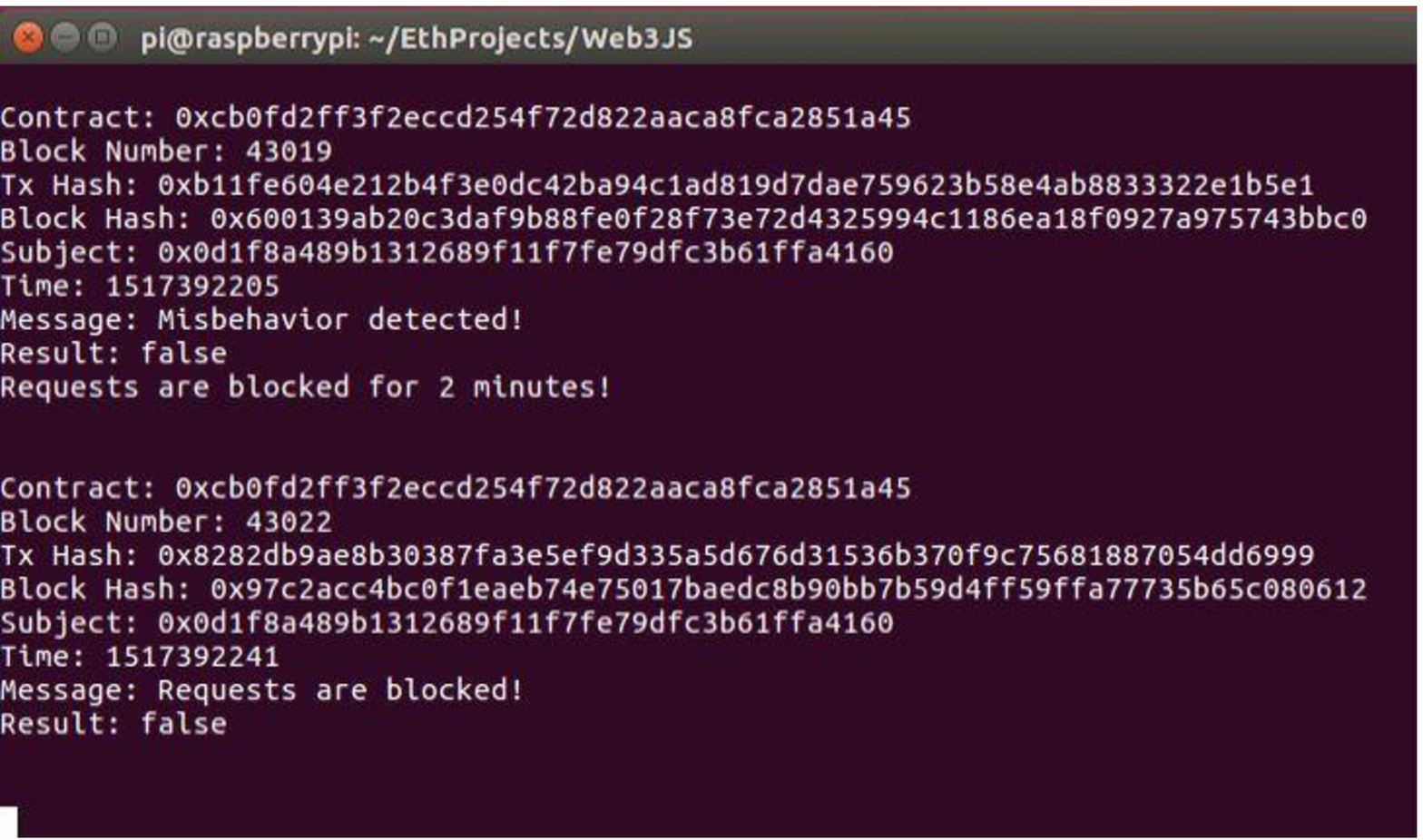}
\label{fig_monitor_2}}
\hfil
\subfloat[Results at the subject.]{\includegraphics[width=3.5in]{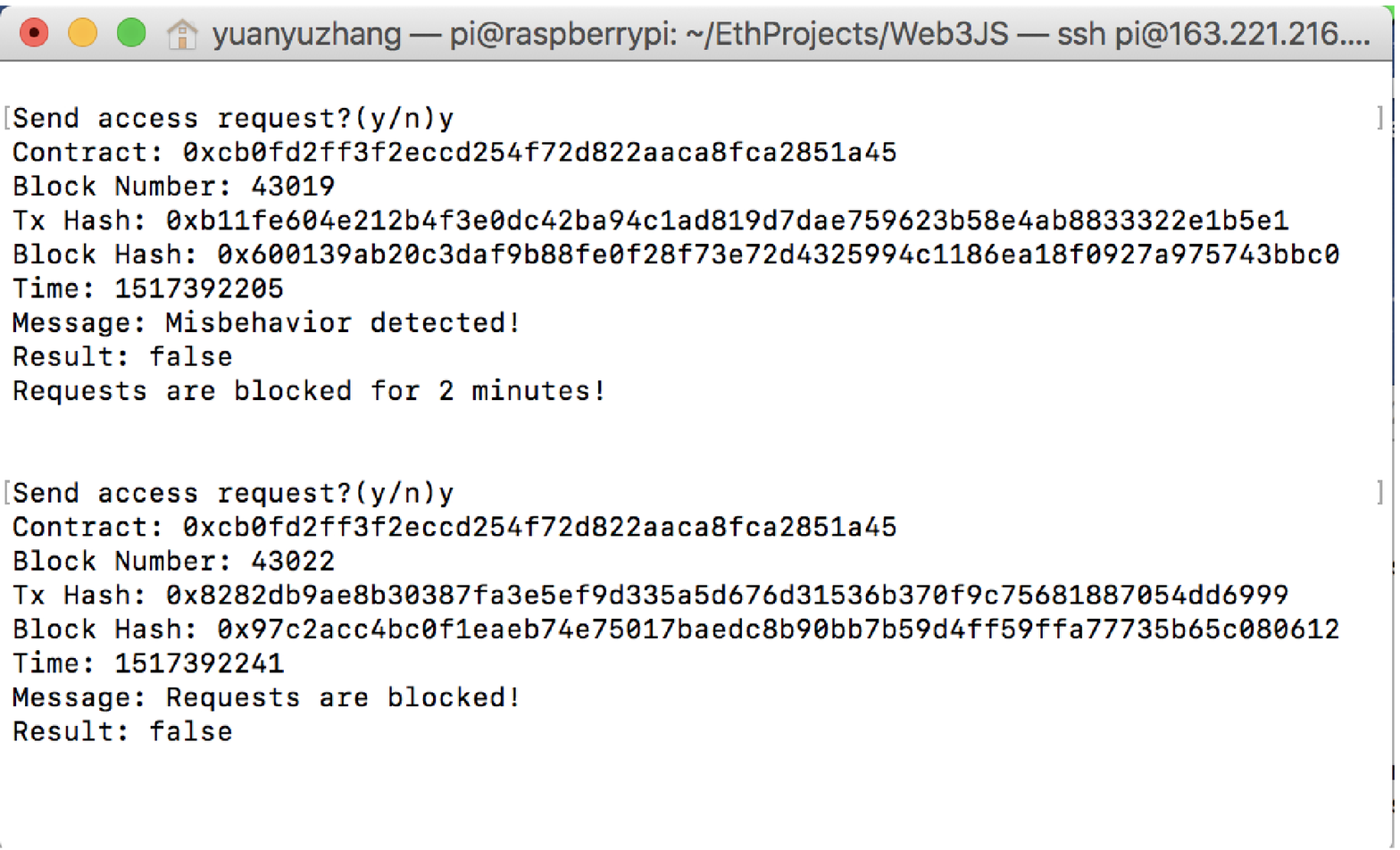}
\label{fig_requester_2}}
\caption{Access results after misbehavior occurring for three times.}
\label{fig_result_2}
\end{figure}

\begin{figure}[!t]
\centering
\subfloat[Results at the object.]{\includegraphics[width=3.5in]{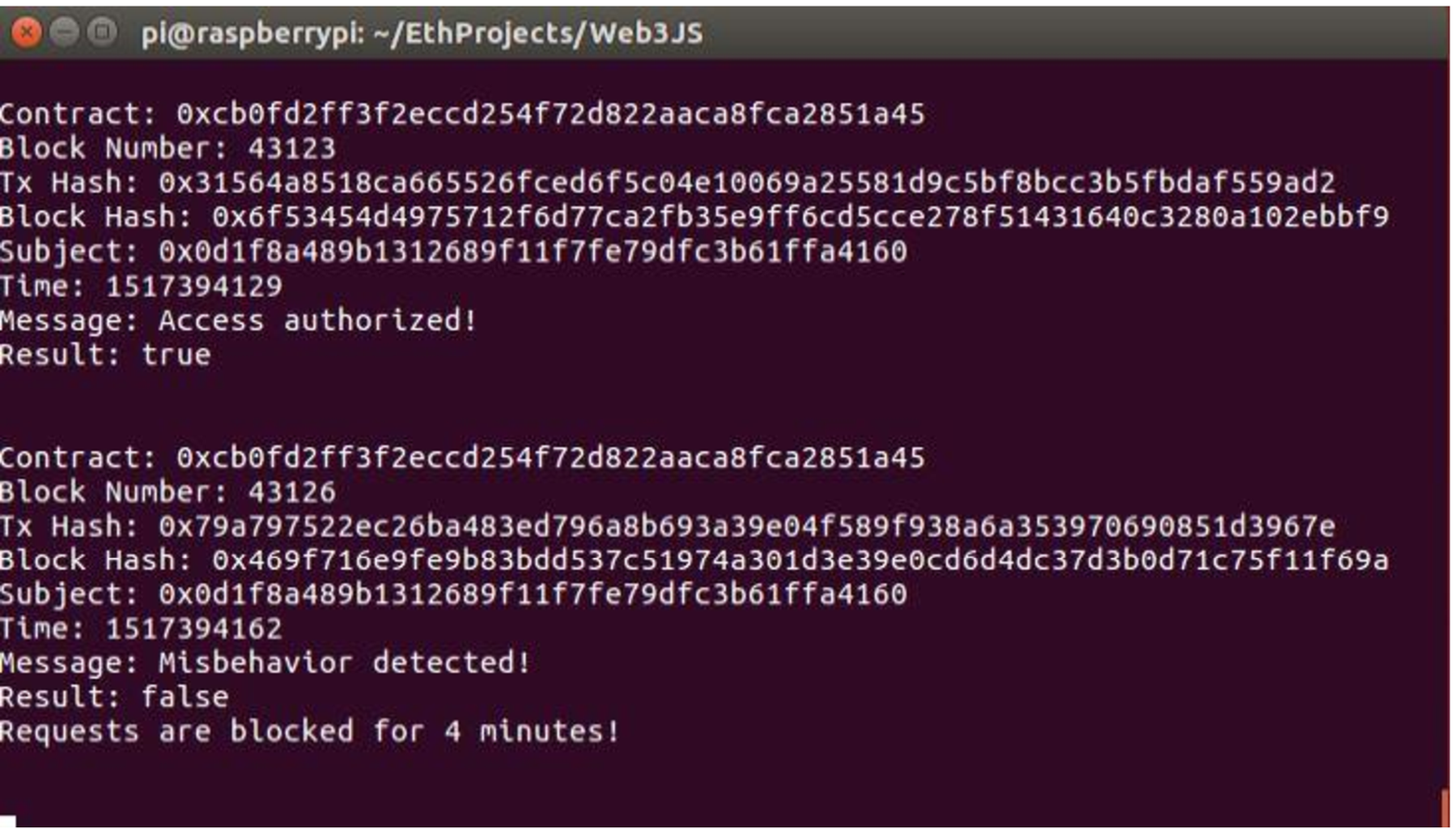}
\label{fig_monitor_3}}
\hfil
\subfloat[Results at the subject.]{\includegraphics[width=3.5in]{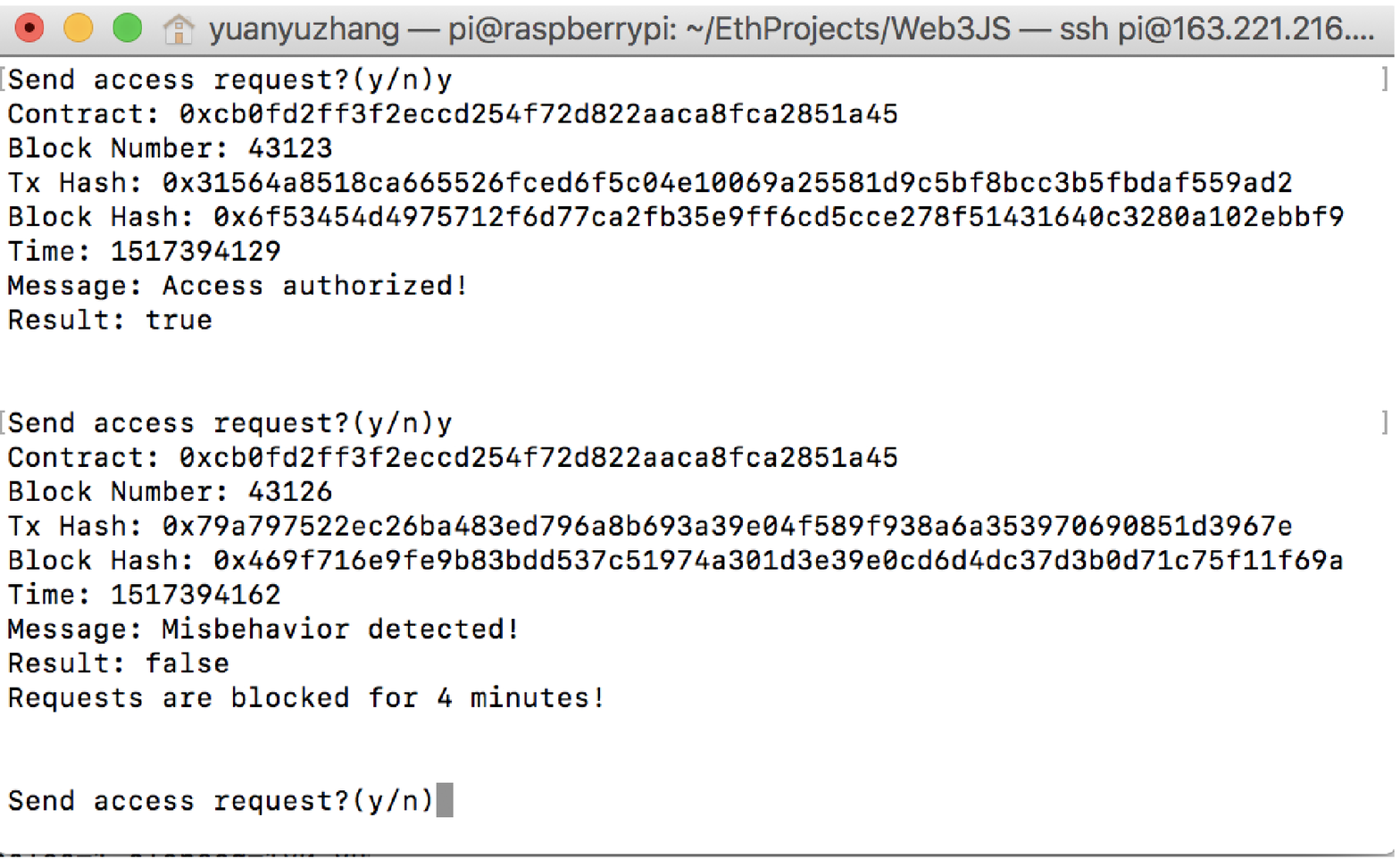}
\label{fig_requester_3}}
\caption{Access results after misbehavior occurring for six times.}
\label{fig_result_3}
\end{figure}

Our source code for the ACC, JC, RC and JavaScripts of the case study is now available at \cite{Sc-IoT}. Based on the code, the hardware and software, we conducted experiments to show the feasibility of the framework for access control. We added a policy to the ACC with \emph{minInterval} set to 100 seconds and \emph{threshold} set to 2. We also set the \emph{base} and \emph{interval} in the JC to $2$ and $3$, respectively.
\begin{figure}[!t]
\centering
\subfloat[Results at the object.]{\includegraphics[width=3.5in]{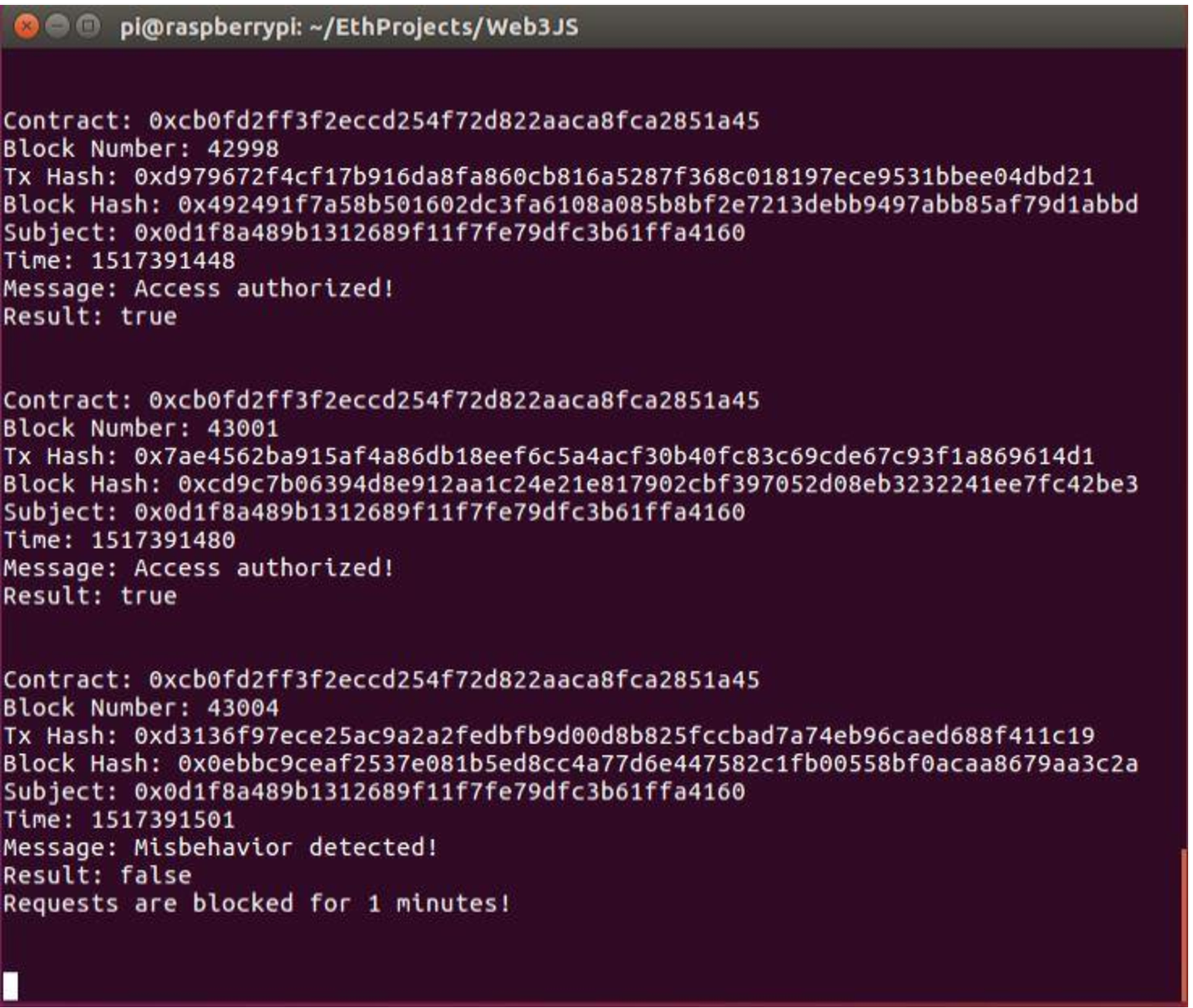}
\label{fig_monitor_1}}
\hfil
\subfloat[Results at the subject.]{\includegraphics[width=3.5in]{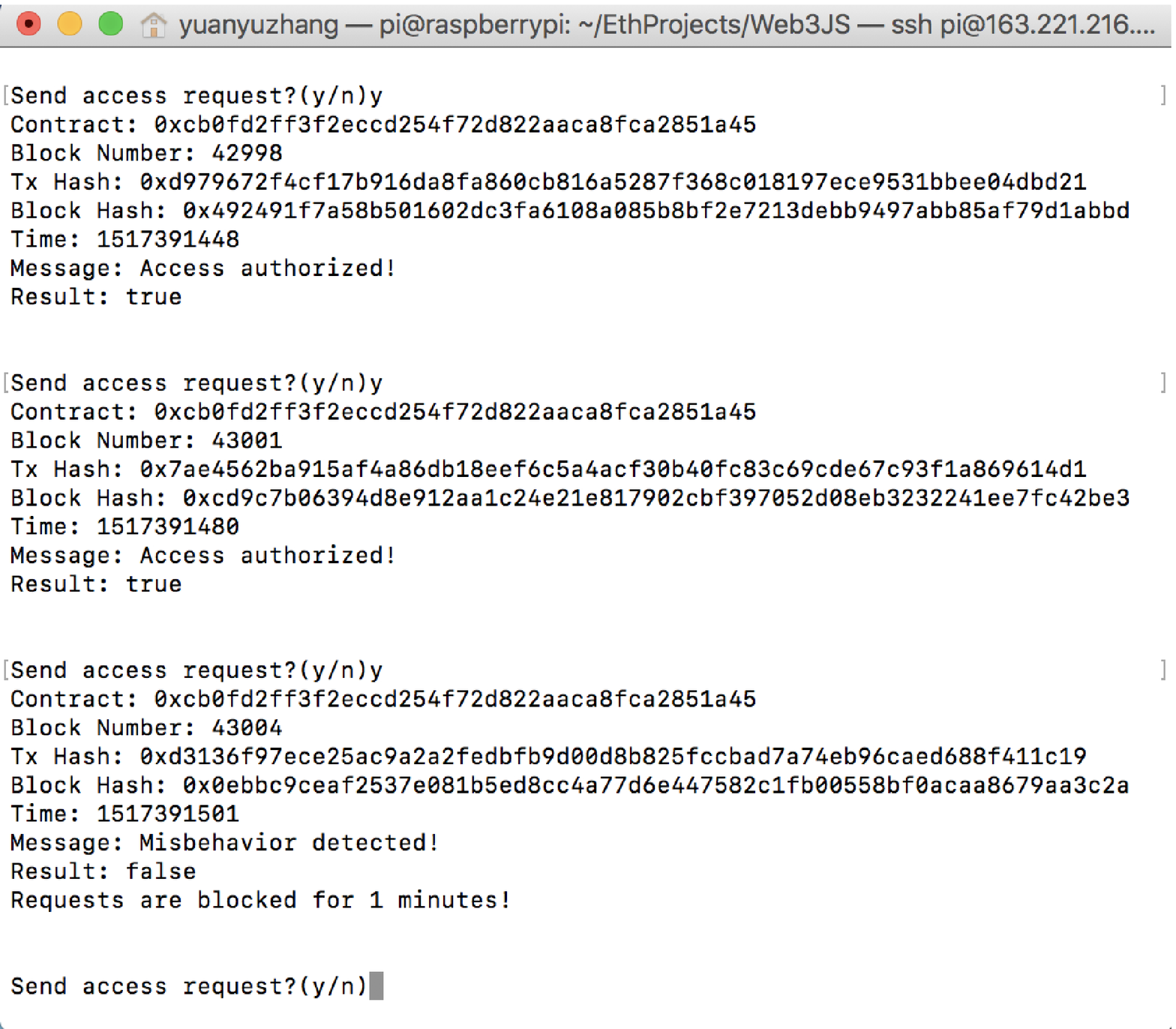}
\label{fig_requester_1}}
\caption{Access results after misbehavior occurring once.}
\label{fig_result_1}
\end{figure}
Fig. \ref{fig_result_1} shows the access results displayed by the JavaScripts at the object (Fig. \ref{fig_monitor_1}) and subject (Fig. \ref{fig_requester_1}), when the subject exhibited the misbehavior for the first time. Fig. \ref{fig_result_2} and Fig. \ref{fig_result_3} show the access results, when the subject exhibited the misbehavior for three times and six times, respectively. We can see that the request of the subject is blocked for $1$, $2$ and $4$ minutes in Fig. \ref{fig_result_1}, Fig. \ref{fig_result_2} and Fig. \ref{fig_result_3}, respectively, which is consistent with the penalty determining equation in (\ref{eqn_penalty}).

\section{Conclusions}\label{sec_con}
This paper investigated the access control issue in the IoT, for which we proposed a smart contract-based framework to implement distributed and trustworthy access control. The framework includes multiple access control contracts (ACCs) for access control between multiple subject-object pairs in the system, one judge contract (JC) for judging the misbehavior of the subjects during the access control, and one register contract (RC) for managing the ACCs and JC. A case study was also provided for the access control in a IoT system with one desktop computer, one laptop and two Raspberry Pi single-board computers. The case study demonstrated the feasibility of the proposed framework in achieving distributed and trustworthy access control for the IoT.

\bibliographystyle{IEEEtran}
\bibliography{MyReferences}

\end{document}